\documentclass{article}

\usepackage{arxiv}

\usepackage[utf8]{inputenc} % allow utf-8 input
\usepackage[T1]{fontenc}    % use 8-bit T1 fonts
\usepackage{hyperref}       % hyperlinks
\usepackage{url}            % simple URL typesetting
\usepackage{booktabs}       % professional-quality tables
\usepackage{amsfonts}       % blackboard math symbols
\usepackage{nicefrac}       % compact symbols for 1/2, etc.
\usepackage{microtype}      % microtypography
\usepackage{lipsum}		% Can be removed after putting your text content
\usepackage{graphicx,url}
\usepackage[sort&compress]{natbib}
\usepackage{doi}
\usepackage{multirow}
\usepackage{units}
\usepackage{color, colortbl}
\usepackage{caption}
\usepackage[list=true]{subcaption}
\usepackage{enumitem}
\usepackage{comment}
\usepackage{float}
\usepackage{fdsymbol}
\usepackage{color, colortbl}
\usepackage{graphicx}
\usepackage[table,xcdraw]{xcolor}

\title{AIRCADE: an Anechoic and IR Convolution-based Auralization Data-compilation Ensemble}

\author{ \href{https://orcid.org/0009-0002-9822-9761}{\includegraphics[scale=0.06]{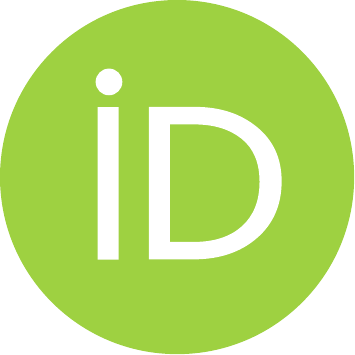}\hspace{1mm}Túlio Chiodi} \\
        School of Acoustical Engineering\\
	Federal University of Santa Maria, Brazil \\
	\texttt{tulio.chiodi@eac.ufsm.br} \\
	\And
	\href{https://orcid.org/0000-0002-3989-7105}{\includegraphics[scale=0.06]{orcid.pdf}\hspace{1mm}Arthur Nicholas dos Santos} \\
	School of Electrical and Computer Engineering\\
	University of Campinas, Brazil \\
	\texttt{a264372@dac.unicamp.br} \\
	\And
	\href{https://orcid.org/0009-0005-2028-9416}{\includegraphics[scale=0.06]{orcid.pdf}\hspace{1mm}Pedro Martins} \\
	School of Electrical and Computer Engineering\\
	University of Campinas, Brazil \\
	\texttt{p137267@dac.unicamp.br } \\
	\And
	\href{https://orcid.org/0000-0002-2246-4450}{\includegraphics[scale=0.06]{orcid.pdf}\hspace{1mm}Bruno Sanches Masiero} \\
	School of Electrical and Computer Engineering\\
	University of Campinas, Brazil \\
	\texttt{masiero@unicamp.br} \\
}

\renewcommand{\shorttitle}{An Anechoic and IR Convolution-based Auralization Data-compilation Ensemble}

\hypersetup{
pdftitle={AIRCADE: an Anechoic and IR Convolution-based Auralization Data-compilation Ensemble},
pdfsubject={},
pdfauthor={},
pdfkeywords={},
}

\begin{document}
\maketitle
%\tableofcontents

\begin{abstract}

In this paper\footnote[1]{This work was partially supported by the São Paulo Research Foundation (FAPESP), grants \mbox{$\#$2017/08120-6} and $\#$2019/22795-1.}, we introduce a data-compilation ensemble, primarily intended to serve as a resource for researchers in the field of dereverberation, particularly for data-driven approaches. It comprises speech and song samples, together with acoustic guitar sounds, with original annotations pertinent to emotion recognition and Music Information Retrieval (MIR). Moreover, it includes a selection of impulse response (IR) samples with varying Reverberation Time (RT) values, providing a wide range of conditions for evaluation. This data-compilation can be used together with provided Python scripts, for generating auralized data ensembles in different sizes: \textit{tiny}, \textit{small}, \textit{medium} and \textit{large}. Additionally, the provided metadata annotations also allow for further analysis and investigation of the performance of dereverberation algorithms under different conditions. All data is licensed under Creative Commons Attribution 4.0 International License.
    
\end{abstract}

% keywords can be removed

\section{Introduction}

Reverberation is the persistence of sound in a space after the source ceases to emit it. It is mainly caused by reflections of the sound waves off the surfaces within the space and gradually dissipates over time. Since the characteristics of the space in which reverberation occurs can influence the duration and intensity of this phenomenon, the Reverberation Time (RT) is considered an important aspect of room acoustics~(\cite{beranek2004concert, moylan2010understanding}).

In the context of entertainment, reverberation can affect the Human Auditory System (HAS) in positive ways, since some degree of it can help to enhance the perceived loudness and richness of sounds. This occurs because the multiple reflections can create a sensation of spaciousness and immersion, which can be aesthetically pleasing, particularly in music. On the other hand, in the context of communications, excessive reverberation can affect the quality and intelligibility of speech because the multiple reflections can create a "\textit{smearing}" effect that can make it difficult to distinguish individual sounds and syllables. Moreover, the reduction of the overall Signal-to-Noise Ratio (SNR) of a given sound source can also mask quiet sounds, making it harder to understand them~(\cite{gelfand2004hearing, lyon2017human}).

Therefore, the effects of reverberation on the HAS can vary widely, depending on the duration and intensity of the phenomenon, the frequency content of the sound source, the individual characteristics of the listener's hearing, etc. In general, it is desirable to optimize the amount of reverberation in a given acoustic scenario to ensure the best possible listening experience for the intended audience. In this situation, dereverberation is a process that can be used to reduce or remove the effects of excessive reverberation from an audio signal. This is typically used in situations where the audio signal has already been recorded in a reverberant environment, and the resulting reverberation is unwanted or detrimental to the quality of the recorded audio~(\cite{naylor2010speech}).

In recent years, data-driven dereverberation methods have become increasingly popular due to their ability to learn complex mappings between the anechoic and reverberant signals. While showing promising results, these methods also have a wide range of applicability, including speech recognition, speaker verification, music production, etc~(\cite{xu2014regression, hershey2016deep}).

Still, an important challenge for these methods is the need for large amounts of training data, with paired clean and reverberant audio signals for training, which, ideally, should cover a wide range of acoustic environments, microphone types, and speaker characteristics, to ensure that a model will generalize well to new scenarios. Examples of dereverberation datasets include the REVERB challenge dataset~(\cite{kinoshita2013reverb}), BUT Speech@FIT Reverb Database~(\cite{szoke2019building}), and VoiceBank-SLR~(\cite{fu2022metricgan}), among others.

However, most of these datasets carry some limitations, such as narrow-band anechoic data, i.e., only speech is considered as a signal of interest, and absence of Impulse Response (IR) data, i.e., usually only anechoic and reverberant data are paired for supervised training. Hence, in this paper, we introduce a new data-compilation ensemble, primarily intended for training data-driven dereverberation models capable of dealing with full bandwidth audio signals, e.g., speech, song, music etc. We offer pairs of natural anechoic and IR data, compiled from datasets licensed under \href{https://creativecommons.org/licenses/by/4.0/legalcode}{Creative Commons Attribution 4.0 International License}, together with Python scripts for convolution-based auralization, under the hypothesis that these ensembles could serve as better training and evaluation tools for such algorithms.

The remainder of this paper is organized as follows: Section \ref{sec2} details the methodology used for selecting the anechoic and IR data, and then synthesizing the auralized data. Section \ref{sec3} describes the resulting auralized data and their annotations. Finally, Section \ref{sec4} presents an overall discussion of our obtained results, together with some pertinent considerations to conclude our study.

\section{Methods}\label{sec2}

To simulate or recreate an acoustic environment, such as a concert hall or a recording studio, using computer algorithms and specialized software, one can resort to a well known technique called auralization. It involves measurements or simulations of the space's physical properties to mimic its acoustic characteristics. To proceed accordingly, there are various techniques available, such as acoustic ray tracing and convolution-based auralization~(\cite{kleiner1993auralization}).

From a signal processing point of view, the convolution is a mathematical operation that describes the interaction between two signals. In the context of auralization, convolution can be used to simulate the effects of an acoustic environment on an audio signal. The basic idea is to convolve an audio signal with an IR that describes the acoustic characteristics of a room or space. The resultant output produces a new signal that represents the original audio signal after it has been modified by the acoustic characteristics of the room~(\cite{allen1979image, oppenheim1997signals}).

In the context of data-driven dereverberation methods, the choice for either natural, synthetic or \textit{joint} datasets, i.e., those obtained by processing combinations of natural recordings with synthetic sounds, directly impacts \textit{in-the-wild} applicability. This occurs because, when models are trained on mostly synthetic data, they usually don’t generalize well for real-world scenarios. However, the
majority of recent studies use joint combinations, e.g., by convolving anechoic data with naturally recorded or synthesized IR data etc, perhaps because it would be too cumbersome to naturally acquire all the necessary data~(\cite{dosSantos2022}). Hence our dataset attempts to reach a balance between these features, compiling real anechoic and IR data, and then synthetically producing auralized data ensembles by means of convolution.

\subsection{Anechoic data}

Trying to cover a wide range of full-bandwidth audio signals, we chose to compile signals of interest from three different categories: speech, song, and musical instruments. 

Since RAVDESS~(\cite{livingstone2018ryerson}) is a well-known dataset with subsets of emotional speech and song, it was chosen to cover the first two aforementioned categories. Its speech-only portion comprises $1,440$ samples performed by $24$ actors ($12$ male and $12$ female), vocalizing $2$ lexically-matched statements ("\textit{kids are talking by the door}" and "\textit{dogs are sitting by the door}") in a "neutral" North American accent. Each expression is pronounced at $2$ different levels of emotional intensity (normal and strong), with an additional neutral expression, resulting in a total amount of $60$ trials per actor. Speech emotions include calm, happy, sad, angry, fearful, surprise, and disgust. Its song-only portion is quite similar and comprises $1,012$ samples performed by $23$ actors, singing the same $2$ lexically-matched statements. Song emotions include only neutral, calm, happy, sad, angry, and fearful expressions. The original sample rate is fixed at 48\,kHz, and Figures \ref{fig:dur} (a) and (b) illustrate histograms with the original duration of files in each RAVDESS subset that we used.

Since the acoustic guitar is a popular musical instrument,  for a variety of reasons, including its ability to produce polyphonic sound and its musical versatility, we chose to use a subset from GuitarSet~(\cite{xi2018guitarset}), referred to as \textit{audio\_mono\--mic}. It comprises $360$ samples performed by $6$ musicians, playing $30$ twelve to sixteen bar excerpts from lead-sheets in a variety of keys, tempos, and musical genres. Recording was performed using a Neumann U87 condenser microphone, placed at approximately 30\,cm in front of the 18th fret of the guitar. The original sample rate is fixed at 44.1\,kHz, and Figure \ref{fig:dur} (c) illustrates a histogram with the original duration of files in this GuitarSet subset.

\subsubsection{Anechoic data processing}

Considering the great difference between the duration of signals in RAVDESS and GuitarSet, we chose to split the samples in GuitarSet into segments of smaller duration, fixed at 5\,s, resulting in $2,004$ different samples. Another reason behind this decision is that when choosing the length of anechoic data, it is important to strike a balance between the computational cost of the convolution operation and the length of the segments. If they are too short, the resulting audio signals may not capture the full extent of the room's acoustics, and if the segments are too long, the convolution operation may become too computationally intensive.

Moreover, since the sample rates of RAVDESS and GuitarSet are different, we also chose to up-sample the GuitarSet segments to 48\,kHz, thus standardizing this value for all files in our dataset.

\subsection{IR data}

\begin{figure}[h!]
\centering
\includegraphics[width=.98\textwidth]{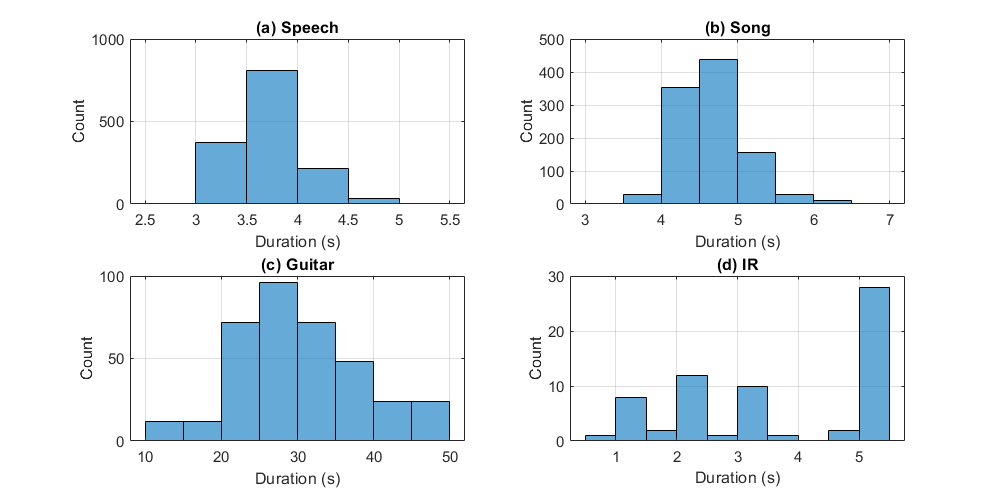}
\caption{Original duration of files in the (a) speech, (b) song, (c) guitar, and (d) IR subsets of our compilation.}
\label{fig:dur}
\end{figure}

IR data was curated from the~\href{https://www.openair.hosted.york.ac.uk/}{\texttt{Open Acoustic Impulse Response (Open AIR) Library}}, which is an online database of Acoustic Impulse Response (AIR) data. Since the original metadata in this database provides information about space category, IR duration, etc., the samples were chosen in order to have a balance between a selected variety of open and enclosed spaces, with IRs in the range of a few milliseconds up to  5\,s. Figure \ref{fig:dur} (d) illustrates a histogram with the original duration of the selected IR data. Altogether, $65$ IRs were chosen.

\subsubsection{IR data processing}

The selected IR data was found in a variety of formats, e.g., B-format, MS Stereo, mono, etc. And at sample rates varying from 44.1\,kHz to 96\,kHz. Since all the anechoic data was compiled in mono format at 48\,kHz, the selected IR data was first converted to mono, then normalized to prevent files from clipping, and finally, each IR was either down- or up-sampled to 48\,kHz. 

\section{Results}\label{sec3}

The processed anechoic and IR data is hosted at \href{https://zenodo.org/record/7818761#.ZDc1snbMJPY}{Zenodo}, with an approximate total file size of 1.3\,GB. For simplicity, all samples in our data-compilation were renamed, e.g., $guitar\_0000$, $rir\_0000$, $song\_0000$, $speech\_0000$, and so on. To synthesize different versions of the auralized data ensemble, the reader is referred to \href{https://github.com/TulioChiodi/AIRCADE}{GitHub}, where Python scripts are available for downloading the base data-compilation and synthesizing a chosen version of the auralized data ensemble. Table \ref{tab:results} illustrates the differences between all versions, detailing the number of song, speech, guitar, IR and auralized samples in each one, together with their respective total file size and duration.

\begin{table}[h!]
\caption{Number of anechoic, IR and resultant auralized data samples, together with their respective total duration and file size for each ensemble version.}
\label{tab:results}
\centering{
\resizebox{.78\textwidth}{!}{%
\begin{tabular}{c|c|c|c|c}
                           & \textbf{Tiny} & \textbf{Small} & \textbf{Medium} & \textbf{Large} \\ \hline
\rowcolor[HTML]{EFEFEF} 
\textbf{Song samples}      & 100           & 500            & 1,012            & 1,012           \\ \hline
\textbf{Speech samples}    & 100           & 500            & 1,012            & 1,440           \\ \hline
\rowcolor[HTML]{EFEFEF} 
\textbf{Guitar samples}    & 100           & 500            & 1,012            & 2,004           \\ \hline
\textbf{IR samples}        & 5             & 9              & 33              & 65             \\ \hline
\rowcolor[HTML]{EFEFEF} 
\textbf{Auralized samples} & 1,500         & 13,500         & 100,188         & 289,640        \\ \hline
\textbf{Total duration}    & 3.2\,h        & 30.41\,h       & 221.77\,h       & 658.08\,h      \\ \hline
\rowcolor[HTML]{EFEFEF} 
\textbf{Total file size}   & 1.1\,GB       & 10.5\,GB       & 76.6\,GB        & 227.5\,GB     
\end{tabular}%
}
}
\end{table}

\subsection{Anechoic data annotation}

Since the ancehoic data in our compilation is comprised of speech and song samples with original annotations pertinent to emotion recognition, together with acoustic guitar sounds with original annotations pertinent to Music Information Retrieval (MIR), we provide metadata which can be used to trace back each sample to its original annotations. This is done because we do not intend for this dataset to be limited to dereverberation tasks only, but also to be used for applications such as emotion recognition and MIR in more challenging scenarios, i.e., in the presence of convolutive noise. 

\subsection{IR data annotation}

\begin{figure}[h!]
\centering
\includegraphics[width=.98\textwidth]{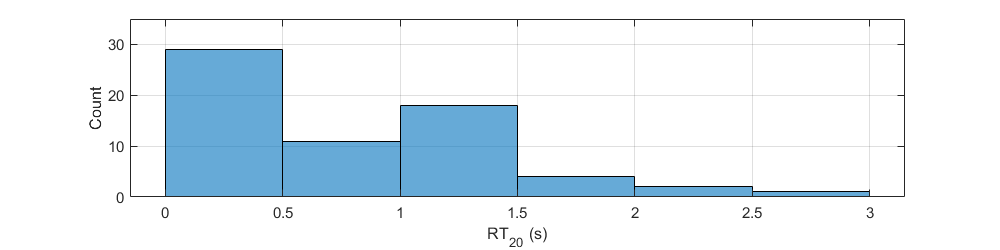}
\caption{Extracted $RT_{20}$ values from the selected the IR data.}
\label{fig:rt20}
\end{figure}

Since the computational effort in dereverberation tasks is highly intertwined with the RT values of IR data, $RT_{20}$ values were extracted from each IR sample using ~\href{https://www.ita-toolbox.org/}{\texttt{ITA Toolbox}}. Figure \ref{fig:rt20} illustrates a histogram with the extracted $RT_{20}$ values from the selected the IR data.

\section{Discussion and Conclusion}\label{sec4}

Overall, the data-compilation ensemble presented in this work provides a diverse and comprehensive set of acoustic scenes for use in dereverberation tasks, as well as some other audio signal processing applications, such as emotion recognition and MIR. By combining different types of signals of interest, including speech, song, and acoustic guitar sounds, with a variety of IRs, we provide a challenging dataset for researchers working on dereverberation and related fields.

The dataset is available in different sizes, from a tiny version with limited data, to a large version with almost $300,000$ samples, allowing users to choose the most suitable version for their specific research needs. The dataset also includes metadata that can be used to trace back each sample to its original annotations, facilitating the use of the dataset for tasks such as emotion recognition and MIR. 

We hope that this dataset will be useful for researchers working on dereverberation and related fields, and we encourage its use in future research. We also believe that the diversity and variability of the dataset can facilitate the development of more robust and generalizable algorithms for dereverberation and other audio signal processing tasks.

\bibliographystyle{unsrtnat}
\bibliography{template}  

\end{document}